\documentclass[12pt,a4paper]{iopart}

\usepackage{iopams}							% packets for math
\usepackage{setstack}						% for e.g. subscripts in summation
\usepackage[english]{babel}			% encoding and language
\usepackage{graphicx}						% packets for pictures
\usepackage{hyperref}						% links in pdf files
\usepackage{cite}								% citation
\usepackage[dvipsnames]{xcolor}

\usepackage{bm}
\newcommand{\PT}{\texorpdfstring{$\mathcal{PT}$}{PT}}
\newcommand{\N}{N_{\mathrm{tot}}}
\newcommand{\imag}{\mathrm {Im} \,}
\newcommand{\real}{\mathrm {Re} \,}

\begin{document}

% -----------------------------------------------------------------------------
% title
\title[Using mixed many-body particle states]{Using mixed many-body particle states to generate exact $\mathcal{PT}$-symmetry in a time-dependent four-well system}
\date{\today}
\author{Tina Mathea$^1$, Dennis Dast$^1$, Daniel Dizdarevic$^1$, Holger Cartarius$^2$, J\"org Main$^1$ and G\"unter Wunner$^1$}
\address{$^1$ Institut f\"ur Theoretische Physik 1, Universit\"at Stuttgart, 70550 Stuttgart, Germany}
\address{$^2$ Physik und ihre Didaktik, Universit\"at Stuttgart, 70550 Stuttgart, Germany}
\ead{Tina.Mathea@itp1.uni-stuttgart.de}

% -----------------------------------------------------------------------------
\begin{abstract}
Bose-Einstein condensates with balanced gain and loss in a double-well potential have been shown to exhibit {\PT}-symmetric states. As proposed by Kreibich \etal [Phys.\ Rev.~A \textbf{87}, 051601(R) (2013)], in the mean-field limit the dynamical behaviour of this system, especially that of the {\PT}-symmetric states, can be simulated by embedding it into a Hermitian four-well system with time-dependent parameters.
In this paper we go beyond the mean-field approximation and investigate many-body effects in this system, which are in lowest order described by the single-particle density matrix. The conditions for {\PT} symmetry in the single-particle density matrix cannot be completely fulfilled by using pure initial states. 
Here we show that it is mathematically possible to achieve exact {\PT} symmetry in the four-well many-body system in the sense of the dynamical behaviour of the single-particle density matrix. In contrast to previous work, for this purpose, we use mixed initial states fulfilling certain constraints and use them to calculate the dynamics.
\end{abstract}

\pacs{03.75.Kk, 03.65.Aa, 11.30.Er}

% -----------------------------------------------------------------------------
\section{Introduction}
\label{sec:introduction}

A Hamiltonian $\hat{H}$ commuting with the product of the parity operator $\mathcal{P}$ and the time reversal operator $\mathcal{T}$, i.e.\ $[\hat{H}, \mathcal{PT}] = 0$, provides unusual properties, therefore the interest in such Hamiltonians has tremendously grown in recent years. Although these Hamiltonians generally are non-Hermitian, they can have stationary states and real eigenvalue spectra \cite{Bender1998, Bender1999}. Since Bender and Boettcher introduced this kind of non-Hermitian quantum mechanics in 1998, there has been much progress in this field \cite{Bender2007, Konotop2016}.

First experimental realizations of {\PT} symmetry succeeded in optical systems \cite{Guo2009,Rueter2010, Klaiman2008, Musslimani2008, Musslimani2010, Peng2014}. Since the formalism of {\PT} symmetry can be used to effectively describe an open quantum system with balanced gain and loss \cite{Dast2014}, Klaiman \etal \cite{Klaiman2008} proposed a Bose-Einstein condensate (BEC) in a laser generated optical double-well potential to be an appropriate candidate to realize {\PT} symmetry in a genuine quantum system. Here, coherent in- and out-coupling produces a {\PT}-symmetric state \cite{Graefe2012}.

The double-well system has been investigated in detail within the limits of a mean-field (MF) approximation by using a Gross-Pitaevskii equation (GPE) \cite{Graefe2008, Graefe2008a, Graefe2010, Haag2012, Cartarius2012, Dast2013, Dast2013a, Dizdarevic2015} and a two-mode approximation \cite{Graefe2012}. In- and out-coupling of particles is effectively described by using complex potentials \cite{Graefe2012}. This leads to an effective and non-Hermitian description of the system.

A possible experimental realization of the two-mode model by constructing a larger Hermitian system, whose parts behave like the open system, was given by Kreibich \etal \cite{Kreibich2013, Kreibich2014, Kreibich2014a}. In their approach the two-mode model is embedded into a larger Hermitian system. The additional wells act as particle reservoirs and are coupled to the other wells by tunnelling processes, which lead to  currents between the wells. Such a larger system provides certain parameters, which can be used to simulate the behaviour of the open system and which, for this purpose, must be chosen time-dependently. In the MF, it is possible to adjust {\PT} symmetry in the inner wells of an optical four-well system \cite{Kreibich2013}, which is a minimal setup of a larger Hermitian system with four real control parameters.

Although the MF description provides a good approximation for large particle numbers and low temperatures \cite{Graefe2008, Graefe2010}, in the system just described many-body effects beyond the MF approximation play an important role \cite{Dast2016a}. Therefore, the question arises whether it is also possible to realize {\PT} symmetry in a full many-body description of the system. In this case, {\PT} symmetry has to be understood in the sense of creating the dynamical behaviour of the {\PT}-symmetric states of the two-mode model in the single-particle matrix of the many-body system.

In a first approach, this has been investigated by Dizdarevic \etal \cite{Dizdarevic2018} in the four-well many-body system introduced by Kreibich \etal \cite{Kreibich2013}. Since they used product states of single-particle states as initial states for the dynamics, {\PT} symmetry could not be realized completely. With the approach of using pure initial states the stationarity of the occupation numbers in the inner wells and of the current between the inner wells could be realized, but the phase correlation did not show the requested behaviour.

In this paper we present a different approach based on the use of
mixed initial states, i.e.\ states which describe an \emph{impure} BEC and cannot be written as products of single-particle states, for the dynamics of the many-body system. We mathematically show that it is possible to realize exact {\PT} symmetry in terms of the single-particle density matrix in the four-well many-body system. We present a method to construct suitable mixed states fulfilling certain constraints and use these states and appropriate time dependences of the control parameters to adjust the whole behaviour of first order expectation values of the two-mode model embedded in the four-well many-body system.

%----------------------------------------------------------------------

This paper is organized as follows: In \sref{sec:section2} we introduce the {\PT}-symmetric two-mode model for BECs in an optical double-well potential with balanced gain and loss. We then recapitulate the main idea of embedding this open quantum system into a Hermitian four-well system with time-dependent control parameters. Since we are interested in how to realize the dynamics of the two-mode model in the four-well many-body system, \sref{sec:section3} deals with the many-body description of bosons in multi-well potentials with a focus on the Bose-Hubbard model with time-dependent control parameters.
In \sref{sec:section4} we develop a method by which the dynamics of the {\PT}-symmetric eigenstates of the two-mode model can be adjusted in the inner wells of the four-mode many-body system by using appropriate mixed initial states. For this purpose, we first derive the time dependences of the control parameters and then we point out a possibility how to construct appropriate initial states. In \sref{sec:section5} we present the numerical results. Finally, \sref{sec:summary} summarizes the contents of this paper and gives a conclusion.

\section{Realization of {\PT} symmetry in the MF and main ideas}
\label{sec:section2}

\subsection{{\PT}-symmetric two-mode model}
\label{sec:section2_1}

For low temperatures and large particle numbers the dynamics of a BEC in an optical double-well with coherent in- and out-coupling of particles can be described by a non-linear Schr\"odinger equation of Gross-Pitaevskii-type \cite{Graefe2012}. If the wells are deep enough and we have balanced gain and loss in the system, a two-mode approximation \cite{Graefe2012} leads to the {\PT}-symmetric GPE,
\begin{equation}
\mathrm{i} \frac{\partial}{\partial t} \left(
\begin{array}{c}
\psi_1 \\
\psi_2
\end{array}\right) =
\left(\begin{array}{cc}
g |\psi_1|^2 + \mathrm i \gamma & -J \\
-J & g |\psi_2|^2 - \mathrm i \gamma
\end{array}\right) \left( \begin{array}{c}
\psi_1 \\
\psi_2
\end{array}\right).
\label{eq:GPE_twomode}
\end{equation}

In this open quantum system gain and loss of particles is modelled by complex potentials $\pm \mathrm i \gamma$ (see \fref{fig:figure1a_1b}(a)) with the coupling strength $\gamma$ describing the in- and outflux of particles. The tunnelling of bosons from well 1 to well 2 is described by the tunnelling rate $J>0$, which couples the two rows in \eref{eq:GPE_twomode}. In the diagonal elements of the Hamiltonian, $g$ denotes the macroscopic contact interaction and is connected with the mass $m$ of the particles and the s-wave scattering length $a$ via $g = 4\pi \hbar^2 a/m$. Note that we set $\hbar=1$ in what follows.

The MF state of the two-mode system \eref{eq:GPE_twomode} is given by a vector $\bm{\psi} = (\psi_1,\psi_2)^T$, where $\psi_i$ with $i \in \{1,2 \}$ denotes the MF wave function in well $i$ and the number of particles in this well is given by $n_i = |\psi_i|^2$.

In the two-mode model \eref{eq:GPE_twomode} stationary solutions exist \cite{Graefe2012}. The {\PT}-symmetric eigenstate lowest in energy, i.e.\ the ground state, is given by \cite{Graefe2012}
\begin{equation}
\left(
\begin{array}{c}
 \psi_1 \\
\psi_2
\end{array} \right) = \left(
\begin{array}{c}
\sqrt{n} \e^{\mathrm i \varphi} \\
\sqrt{n} \e^{-\mathrm i \varphi}
\end{array} \right)
\label{eq:eigen}
\end{equation}
with the phase
\begin{equation}
\varphi = -\frac{1}{2} \arcsin \left( \frac{\gamma}{J} \right)
\end{equation}
and $n$ the number of particles in well 1 and 2, where $\gamma \leq J$. For the dynamics of the eigenstate \eref{eq:eigen} given by \eref{eq:GPE_twomode} one finds the occupation numbers $n_i(t) = n$ to be constant in time and the reduced current density $\tilde{j}_{12}(t)= -\mathrm{i} (\psi_1^* \psi_2 - \psi_2^* \psi_1) = 2 n \gamma/ J$ and the correlation $c_{12}(t) = \psi_1^* \psi_2 + \psi_2^* \psi_1 = 2 n \sqrt{1-\gamma^2/J^2}$ to be stationary as well. These four real quantities characterize the MF dynamics of the two-mode model \eref{eq:GPE_twomode}. We are interested in a method to adjust this dynamics in a four-well many-body system. The four-well system and its MF behaviour are introduced in the following.

\subsection{Embedding in Hermitian four-well system}

\begin{figure}
\includegraphics[width=0.99\textwidth]{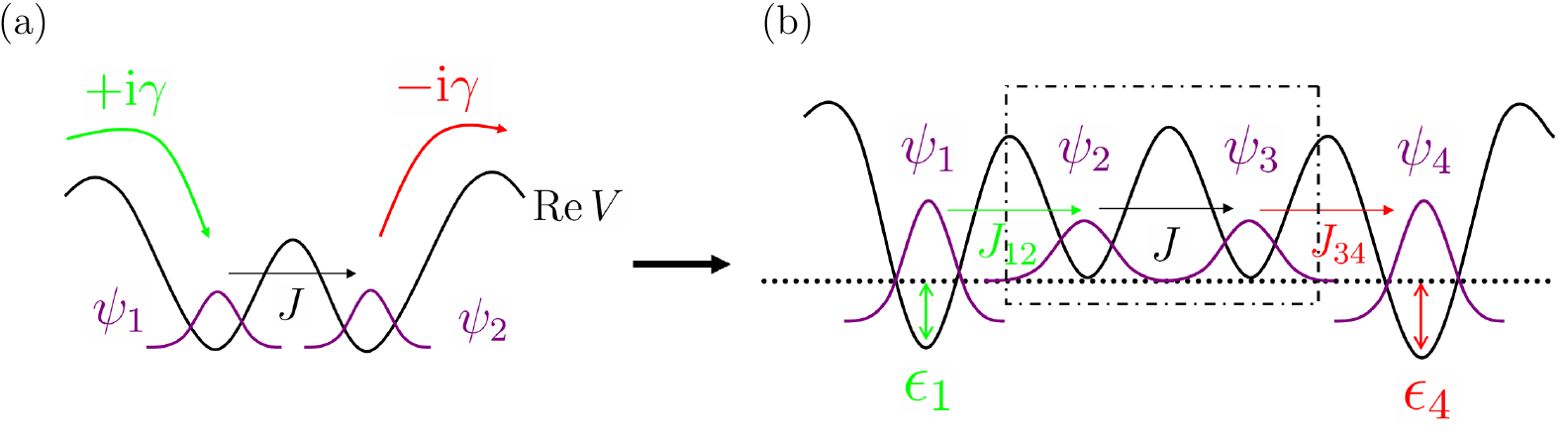}
\caption{\label{fig:figure1a_1b} Embedding a (a) two-mode model into (b) a Hermitian four-well system. The time dependence of the tunnelling rates $J_{12}(t)$ and $J_{34}(t)$ as well as that of the onsite energies $\epsilon_1(t)$ and $\epsilon_4(t)$ can be chosen as required to adjust the dynamics of the two-mode model in the inner wells of the four-well system.}
\end{figure}

The two-mode model \eref{eq:GPE_twomode} is an open quantum system and therefore its Hamiltonian is non-Hermitian. To make an experimental realization of this {\PT}-symmetric but open and non-Hermitian system possible, Kreibich \etal simulated its dynamics by embedding it into a larger but Hermitian system \cite{Kreibich2013, Kreibich2014, Kreibich2014a}. The intention is to control certain parameters, in the following referred to as \emph{control parameters}, in such a way that parts of the Hermitian system behave like the {\PT}-symmetric two-mode model. To observe the dynamics of the {\PT}-symmetric eigenstates \eref{eq:eigen} in parts of the Hermitian system is of special interest. It has been shown \cite{Kreibich2013} that already a four-well system as depicted in \fref{fig:figure1a_1b} is sufficient to realize the dynamics of the two-mode model in the inner wells of the system in the MF. The inner wells are referred to as \emph{system wells}, while the two additional wells act as particle reservoirs, henceforth referred to as \emph{reservoir wells}.

As mentioned above, in the two-mode model \eref{eq:GPE_twomode} we have particle-influx on the left-hand side and outflux on the right-hand side, which both can be described by the imaginary part of the potential. To achieve the same behaviour as the stationary states \eref{eq:eigen} in the system wells, the four-well system has four free and real-valued control parameters, which can be used to adjust the requested dynamics. On the one hand, there are the tunnelling rates $J_{12}(t)$ and $J_{34}(t)$ describing the coupling between the wells by tunnelling processes, which lead to tunnelling currents between the wells. On the other hand, we have the onsite energies $\epsilon_1(t)$ and $\epsilon_4(t)$. It is necessary that these four control parameters are time-dependent \cite{Kreibich2013}. They are varied in time in such a way that the inner wells show the same dynamics as the open system. Note that in the following we choose $\epsilon_2=\epsilon_3=0$ to achieve {\PT} symmetry in the inner wells.

The considerations in \cite{Kreibich2013} only deal with the MF description of the Hermitian four-well system, therefore the time dependences derived are only valid for the MF limit. As shown by Kreibich \etal \cite{Kreibich2013}, it is possible to realize the requested behaviour in the MF limit.

To obtain the dynamics of the {\PT}-symmetric state \eref{eq:eigen} in the system wells it is necessary to choose an appropriate initial state for the equations of motion. The MF state in the four-well system is given by a vector $\bm{\psi} = (\psi_1,\psi_2, \psi_3,\psi_4)^T$, where $\psi_i=\sqrt{n_i} \mathrm{e}^{\mathrm{i}\varphi_i}$ with $i \in \{1,2,3,4 \}$ denotes the MF wave function in well $i$, analogously to the two-mode model. By using the approach of Dizdarevic \etal \cite{Dizdarevic2018} one finds the relations $\varphi_1 - \varphi_2 = -\pi/2$ and $\varphi_4 - \varphi_3 = \pi/2$ for the phases. The initial condition for the MF wave function therefore can be written as \cite{Dizdarevic2018}
\begin{equation}
\left(
\begin{array}{c} \psi_1(0) \\ \psi_2(0) \\ \psi_3(0) \\ \psi_4(0)
\end{array}
\right) = \left(
\begin{array}{c}
-\rmi \sqrt{n_1(0)} \e^{\rmi \varphi} \\
\sqrt{n(0)} \e^{\rmi \varphi} \\
\sqrt{n(0)} \e^{-\rmi \varphi} \\
\rmi \sqrt{n_4(0)} \e^{-\rmi \varphi}
\end{array}
\right),
\label{eq:psi1psi4}
\end{equation}
where $n_1(0)$ and $n_4(0)$ denote the initial occupation numbers in the reservoir wells, which can be chosen freely. In \sref{sec:section4} the MF coefficients in \eref{eq:psi1psi4} will be used to construct an appropriate initial many-body state for the many-body dynamics.

By using a MF description to calculate the dynamics of the system only single-particle effects can be taken into account, whereas many-body effects are neglected. To consider all effects present in the system a many-body description is required. In this paper we focus on a many-body description of the system with the goal of adjusting the dynamics of the {\PT}-symmetric states \eref{eq:eigen} in the inner wells of the four-well many-body system. For this purpose, the Bose-Hubbard model is introduced.

\section{Many-body description via the Bose-Hubbard model}
\label{sec:section3}

The many-body dynamics of bosons trapped in multi-well potentials can be described by the Bose-Hubbard model \cite{Jaksch1998}. In this section we summarize the methods for describing the four-well many-body system used in this work. The description given here is not limited to four-well systems but in general may be also used to describe bosons in optical lattices or multi-well potentials.

For our purpose time-dependent control parameters are necessary. Therefore we use a dynamical Bose-Hubbard model with time-dependent tunnelling rates $J_{kl}(t)$ with $k,l \in \{1,\dots,4\}$ and time-dependent onsite energies $\epsilon_i(t)$ with $i \in \{1,\dots,4\}$. The Bose-Hubbard Hamiltonian then has the form
\begin{equation}
\label{eq:BHM_Ham}
\hspace{-1.2cm} \hat{H}_{\mathrm{BH}} = - \sum_{\langle m,m' \rangle} J_{m,m'}(t) \hat{a}^\dagger_m \hat{a}_{m'} + \frac{1}{2} \sum_m U_m \hat{a}_m^\dagger \hat{a}_m \left( \hat{a}_m^\dagger \hat{a}_m - 1 \right) + \sum_m \epsilon_m(t) \hat{a}_m^\dagger \hat{a}_m,
\end{equation}
where $\hat{a}_m^{\dagger}$ creates and $\hat{a}_m$ annihilates a boson in well $m$. The first term in \eref{eq:BHM_Ham} describes the tunnelling of bosons between neighbouring lattice sites. Only next-neighbour coupling is taken into account by adding up all nearest neighbours, which is denoted by $\langle \cdot, \cdot \rangle$. The strength of the tunnelling coupling is characterized by the tunnelling rate $J_{kl}$. The onsite interaction between two or more bosons localized on the same lattice site is given by the second term in \eref{eq:BHM_Ham}. Their repulsion is quantified by the microscopic contact interaction strength $U=g/(\N-1)$ with $\N$ being the total particle number in the system. The third term takes into account an energy offset $\epsilon_i$ on the $i$th lattice site, which is caused by an external confinement.

Bosons are indistinguishable and at sufficiently low temperatures they are localized in the lowest Bloch band. Therefore a state of the system is totally characterized by the number of particles in each well, which is why a Fock basis can be used to describe the many-body dynamics. In the case of a four-well system a basis vector $|n_1,n_2,n_3,n_4 \rangle$ is determined by the occupation numbers $n_i$ with $i \in \{1,\dots,4\}$. For a total particle number of $\N$ bosons in a system with $M$ wells the dimension $D$ of the corresponding Hilbert space is given by
\begin{equation}
D(\N, M) =  \left( \begin{array}{c} \N + M - 1 \\ \N  \end{array} \right),
\label{eq:dim}
\end{equation}
i.e.\ for large particle numbers the Hilbert-space grows polynomially with $\N$.

The Bose-Hubbard model with time-dependent parameters as presented is used to describe the many-body dynamics of the system in this paper. To calculate the time evolution of the system the methods described in \cite{Zhang_exakt_diag, Dizdarevic2018} are used.

An alternative formulation of the Bose-Hubbard model is given by the Bogoliubov-Born-Green-Kirkwood-Yvon (BBGKY) hierarchy \cite{AnglinVardi2001a}. The many-body dynamics of the system can equivalently be described by the density operator $\hat{\rho}$. By using the von Neumann equation the dynamics can be rewritten in terms of the elements
\begin{equation}
\sigma_{kl} = \langle \hat{a}_k^\dagger \hat{a}_l \rangle
\label{eq:sigma_kl}
\end{equation}
of the single-particle density matrix, the elements
\begin{equation}
\sigma_{klmn} = \langle \hat{a}_k^\dagger \hat{a}_l \hat{a}_m^\dagger \hat{a}_n \rangle
\label{eq:sigma_klmn}
\end{equation}
of the two-particle density matrix, and in general the elements
\begin{equation}
\sigma_{i_1,\dots,i_n} = \langle \hat{a}^\dagger_{i_1} \hat{a}_{i_1} \dots \hat{a}^\dagger_{i_n} \hat{a}_{i_n} \rangle
\end{equation}
of the $n$-particle density matrices. One obtains the BBGKY hierarchy, which is a coupled system of linear differential equations, of the form
\begin{eqnarray}
\mathrm{i} \frac{\partial}{\partial t} \sigma_{kl}  = f \left(  \sigma_{kl} , \sigma_{klmn} \right), \\
\mathrm{i} \frac{\partial}{\partial t} \sigma_{klmn} = f \left( \sigma_{kl} , \sigma_{klmn} , \sigma_{klmnrs} \right), \\
\vdots \nonumber
\end{eqnarray}

For further purposes we here give the first order of the hierarchy explicitly. It can be written as
\begin{equation}
\mathrm i \frac{\partial}{\partial t} \sigma_{kl} = Z_{kl} - \left( \epsilon_k - \epsilon_l \right) \sigma_{kl}
\label{eq:BBR_first_order}
\end{equation}
with the abbreviation
\begin{eqnarray}
\nonumber
Z_{kl} &= J_{k-1,k} \sigma_{k-1,l} + J_{k+1,k} \sigma_{k+1,l} - J_{l,l-1} \sigma_{k,l-1} - J_{l,l+1} \sigma_{k,l+1} \\
&\quad - U_k \left( \sigma_{kkkl} - \sigma_{kl} \right) + U_l \left(\sigma_{klll} - \sigma_{kl} \right).
\label{eq:Z_kl}
\end{eqnarray}
The equation of motion \eref{eq:BBR_first_order} describes the first order of the system's many-body dynamics. We will use \eref{eq:BBR_first_order} to compare the dynamics of the two-mode model with the four-well many-body system to derive time dependences of the control parameters.

\section{Dynamics of mixed states}
\label{sec:section4}

In this section a method to adjust the dynamics of the {\PT}-symmetric states of the two-mode system in the four-well many-body system is presented. It has been shown \cite{Dizdarevic2018} that pure initial states may not be used to create the requested dynamics of all four quantities  $n_2$, $n_3$, $\real \sigma_{23}$ and $\imag \sigma_{23}$, but only three of them. By using pure initial states for the many-body dynamics only the stationarity of the occupation numbers $n_2$ and $n_3$ and the reduced current density $2 \imag \sigma_{23}$ can be achieved, while the correlation $2 \real \sigma_{23}$ cannot be kept constant.

Here we present a different approach based on the use of \emph{mixed} initial states, i.e.\ states which cannot be expressed as products of single-particle states. Due to the contact interaction in the many-body system a pure state represents not an ideal choice of an initial state. To create the dynamics of the two-mode model in the inner wells of the four-well many-body system two problems have to be solved: First, to find suitable time dependences of the control parameters, and second, to construct appropriate initial states for the many-body dynamics.

\subsection{Time dependence of parameters}
\label{sec:time_dep_of_par}

Our starting point for deriving the time dependence of the control parameters is the two-mode model in the MF described by the GPE \eref{eq:GPE_twomode}. To obtain terms for the time dependences of the control parameters we compare the dynamics of \eref{eq:GPE_twomode} with the first order of the many-body dynamics for the four-well system given by \eref{eq:BBR_first_order}. By doing so, we find four real requirements for the adjustable parameters $J_{12}(t)$, $J_{34}(t)$, $\epsilon_1(t)$ and $\epsilon_4(t)$. Note that in this way we only adjust the dynamics of the first order in the many-body system. A more detailed discussion of this point is given in \sref{sec:section5}.

In the MF limit the elements of the single particle density matrix are given by products of the MF coefficients $\psi_i$, i.e.\ $\sigma_{kl} = \psi_k^* \psi_l$. Therefore their time derivatives read
\begin{equation}
\dot{\sigma}_{kl} = \dot{\psi}_k^* \psi_l + \psi_k^* \dot{\psi}_l.
\end{equation}

The time dependence of the tunnelling rates directly results from a comparison of the dynamics of $\sigma_{22}$ and $\sigma_{33}$, respectively. We demand $\dot{\sigma}_{22}$ and $\dot{\sigma}_{33}$, respectively, to be equal in the two-mode model and in the first order of the four-well many-body system. Comparison of the terms obtained with \eref{eq:GPE_twomode} on the one hand and with \eref{eq:BBR_first_order} on the other hand directly yields the expressions
\numparts
\begin{equation}
J_{12}(t)  = \frac{2 \gamma n_2(t)}{\tilde{j}_{12}(t)}
\label{eq:tunnelJ12}
\end{equation}
and
\begin{equation}
J_{34}(t) = \frac{2 \gamma n_3(t)}{\tilde{j}_{34}(t)}
\label{eq:tunnelJ34}
\end{equation}
\endnumparts
for the tunnelling rates with the reduced density current given by
\begin{equation}
\tilde{j}_{kl} = 2 \imag \sigma_{kl}.
\label{eq:jtilde_manybody}
\end{equation}
Note that in the MF limit \eref{eq:jtilde_manybody} gives exactly the expressions from \sref{sec:section2_1}.

Analogously, by comparing the time derivative of the element $\sigma_{23}$ for the two-mode model given by \eref{eq:GPE_twomode} with the four-well many-body system in the first order given by \eref{eq:BBR_first_order} one finds the complex equation
\begin{equation}
- J_{12}(t) \sigma_{13}(t) + J_{34}(t) \sigma_{24}(t) = 0,
\label{eq:Forderung}
\end{equation}
from which the two onsite energies $\epsilon_1(t)$ and $\epsilon_4(t)$ (see \fref{fig:figure1a_1b}(b)) can be obtained as shown below. From this it is clear that the sum of the terms containing $\sigma_{13}$ and $\sigma_{24}$ must vanish, because, of course, we have no reservoir wells in the two-mode model. Note that the requirement \eref{eq:Forderung} has to be fulfilled at time $t=0$, which will have to be taken into account when choosing suitable initial states.

To guarantee that the requirement \eref{eq:Forderung} is fulfilled at all times $t$ its time derivative must vanish, i.e.
\begin{equation}
- \dot{J}_{12}(t)  \sigma_{13}(t) - J_{12}(t) \dot{\sigma}_{13}(t) +  \dot{J}_{34}(t)  \sigma_{24}(t) + J_{34}(t) \dot{\sigma}_{24}(t) = 0.
\label{eq:Forderung_Zeitableitung}
\end{equation}
By using the dynamics \eref{eq:BBR_first_order} of the many-body system and by separating real and imaginary parts of the resulting equation, a system of two linear equations can be found. It is of the form
\numparts
\begin{equation}
\alpha_r \epsilon_1 + \beta_r \epsilon_4 = \Omega_r,\label{LGS_epsilon_a}
\end{equation}
\begin{equation}
\alpha_i \epsilon_1 + \beta_i \epsilon_4 = \Omega_i.
\label{LGS_epsilon_b}
\end{equation}
\endnumparts
The occurring real coefficients $\alpha_r$, $\alpha_i$, $\beta_r$, $\beta_i$, $\Omega_r$, $\Omega_i$ are given by
\numparts
\begin{equation}
\alpha_r = \frac{1}{2} J_{12} \left( \frac{c_{12} c_{13}}{\tilde{j}_{12}} + \tilde{j}_{13} \right),
\label{eq:alpha_r}
\end{equation}
\begin{equation}
\beta_r = \frac{1}{2} J_{34} \left( \frac{c_{34} c_{24}}{\tilde{j}_{34}} + \tilde{j}_{24} \right),
\label{eq:beta_r}
\end{equation}
\begin{eqnarray}
\nonumber
\Omega_r & =  \frac{1}{2} J_{12} \left( \frac{Y_{22} c_{13}}{2 n_2} + \frac{X_{12} c_{13}}{\tilde{j}_{12}} + \frac{X_{22} \tilde{j}_{13}}{2 n_2} + Y_{13} \right) \\
& - \frac{1}{2} J_{34} \left( \frac{Y_{33} c_{24}}{2 n_3} + \frac{X_{34} c_{24}}{\tilde{j}_{34}} + \frac{X_{33} \tilde{j}_{24}}{2 n_3} + Y_{24} \right),
\label{eq:omega_r}
\end{eqnarray}
\begin{equation}
\alpha_i = \frac{1}{2} J_{12} \left( \frac{c_{12} \tilde{j}_{13}}{\tilde{j}_{12}} - c_{13} \right),
\label{eq:alpha_i}
\end{equation}
\begin{equation}
\beta_i = \frac{1}{2} J_{34} \left( \frac{c_{34} \tilde{j}_{24}}{\tilde{j}_{34}} - c_{24} \right),
\label{eq:beta_i}
\end{equation}
and
\begin{eqnarray}
\nonumber
\Omega_i &= \frac{1}{2} J_{12} \left( -\frac{X_{22} c_{13}}{2 n_2}  + \frac{Y_{22} \tilde{j}_{13}}{2 n_2} + \frac{X_{12} \tilde{j}_{13}}{\tilde{j}_{12}} - X_{13} \right) \\
&- \frac{1}{2} J_{34} \left( - \frac{X_{33} c_{24}}{2 n_3} +\frac{Y_{33} \tilde{j}_{24}}{2 n_3} + \frac{X_{34} \tilde{j}_{24}}{\tilde{j}_{34}} -  X_{24} \right),
\label{eq:omega_i}
\end{eqnarray}
\endnumparts
where we used the correlations $c_{kl}$ defined as
\begin{equation}
c_{kl} = 2 \real \sigma_{kl}
\end{equation}
and the abbreviations
\begin{equation}
X_{kl} = 2 \real Z_{kl}
\end{equation}
and
\begin{equation}
Y_{kl} = 2 \imag Z_{kl}.
\end{equation}
The equations \eref{LGS_epsilon_a} and \eref{LGS_epsilon_b} are a two-dimensional linear system of equations for the parameters $\epsilon_1$ and $\epsilon_4$. In the case of a non-vanishing determinant of the system's coefficient matrix a single unique solution exists, which is given by
\numparts
  \begin{equation}
    \epsilon_1 = \frac{\beta_i \Omega_r - \beta_r \Omega_i}{\alpha_r \beta_i - \beta_r \alpha_i},
    \label{Lsg_epsilon1}
  \end{equation}
  \begin{equation}
    \epsilon_4 = - \frac{ \left( \alpha_i \Omega_r - \alpha_r \Omega_i \right)}{\alpha_r \beta_i - \beta_r \alpha_i}.
    \label{Lsg_epsilon4}
  \end{equation}
\endnumparts
Note that the system of equations \eref{LGS_epsilon_a} and \eref{LGS_epsilon_b} does not have a single unique solution in the limit of pure states but infinitely many solutions. In this case the elements of the single-particle density matrix are of the form $\sigma_{kl} = \psi_k^* \psi_l$, which implies that the determinant $(\alpha_r \beta_i - \beta_r \alpha_i)$ yields zero and the numerators $(\beta_i \Omega_r - \beta_r \Omega_i)$ and $(\alpha_i \Omega_r - \alpha_r \Omega_i)$ as well. Thus, the equations \eref{LGS_epsilon_a} and \eref{LGS_epsilon_b} become dependent. This behaviour already arose in investigations concerning the MF properties of the four-well system of Kreibich \etal \cite{Kreibich2013}, which was the reason for introducing an additional time-dependent degree of freedom $d(t)$. This additional degree of freedom can lead to several difficulties and different possibilities for choosing $d(t)$ were already investigated \cite{Kreibich2014a}.

Here, we focus on the use of mixed initial states to adjust the dynamics of the open system in the many-body system. In contrast to the equations in \cite{Kreibich2013} the time dependences \eref{eq:tunnelJ12}--\eref{eq:tunnelJ34} and \eref{LGS_epsilon_a}--\eref{LGS_epsilon_b} presented here also allow for calculating the dynamics of mixed states, for which the onsite energies are well-defined. In the following we will present a way of how to construct appropriate mixed initial states.

\subsection{Construction of appropriate mixed initial states}

Mixed states have many degrees of freedom. In this section we deal with the problem of how suitable mixed initial states for the many-body system can be constructed such that they show the same single-particle dynamics as the {\PT}-symmetric state \eref{eq:eigen} of the open system. The basic idea is to use the MF state, which is a pure state, and to perturb it to obtain a mixed state.

\paragraph{Constraints}

To achieve the dynamical properties of the {\PT}-symmetric eigenstate \eref{eq:eigen}, the initial state has to fulfill certain constraints, precisely five real-valued equations. Two of them directly result from the comparison of time derivatives of the single-particle density matrices in \sref{sec:time_dep_of_par}. As mentioned there, the requirement \eref{eq:Forderung} has to be fulfilled at $t=0$. From this complex equation the two real conditions
\numparts
\begin{equation}
- J_{12}(0) \real \sigma_{13}(0) + J_{34}(0) \real \sigma_{24}(0) \stackrel{!}{=}  0,
\label{eq:Bed1u2}
\end{equation}
\begin{equation}
- J_{12}(0) \imag \sigma_{13}(0) + J_{34}(0) \imag\sigma_{24}(0) \stackrel{!}{=}  0
\end{equation}
arise.

Due to the fact that we intend to realize the dynamics of the stationary state \eref{eq:eigen} in the first order of the four-mode many-body system, there are three other constraints the resulting initial state has to fulfill. The chosen mixed initial state must have the same initial values of $n_2$, $n_3$, $\tilde{j}_{23}$ and $c_{23}$ as the {\PT}-symmetric state \eref{eq:eigen}.

In the first order of the many-body dynamics in terms of \eref{eq:BBR_first_order} the stationary states \eref{eq:eigen} of the two-mode model are determined by $\sigma_{22}=n_2$, $\sigma_{23}=n_3$, $\imag \sigma_{23} \sim \tilde{j}_{23}$ and $\real \sigma_{23} \sim c_{23}$.
To obtain the desired dynamics, the occupation number in the inner wells has to be equal. For the initial many-body state this yields the condition
\begin{equation}
\real \sigma_{22}(0) \stackrel{!}{=}\real \sigma_{33}(0).
\label{eq:Bed3}
\end{equation}
For the reduced current density $\tilde{j}_{23}$ one finds by using \eref{eq:eigen}
\begin{equation}
\imag \sigma_{23}(0) \stackrel{!}{=} \sqrt{\real \sigma_{22}(0) \real \sigma_{33}(0)} \frac{\gamma}{J}.
\end{equation}
Analogously, for the correlation $c_{23}$ the condition reads
\begin{equation}
\real \sigma_{23}(0) \stackrel{!}{=} \sqrt{\real \sigma_{22}(0) \real \sigma_{33}(0)} \sqrt{1- \frac{\gamma^2}{J^2}}.
\label{eq:Bed4u5}
\end{equation}
\endnumparts

The elements of the single-particle density matrix must fulfill the five real requirements \eref{eq:Bed1u2}--\eref{eq:Bed4u5}. Because a mixed state has a huge number of degrees of freedom, there is not only one state fulfilling \eref{eq:Bed1u2}--\eref{eq:Bed4u5}, but there are infinitely many. Thus, there is not only one way to find an appropriate mixed state. In the following we present one possibility to construct an initial state fulfilling the constraints \eref{eq:Bed1u2}--\eref{eq:Bed4u5} by starting from a pure state, the MF state \eref{eq:psi1psi4}, which we will perturb.

\paragraph{Construction from pure states}

Any state $|\bpsi, \N \rangle$ of a system consisting of $M$ wells can be expressed in a Fock basis and therefore can be written as
\begin{equation}
|\bpsi, \N \rangle  = \sum_{n_1,\dots,n_M}  c_{n_1,\dots,n_M} |n_1, \dots, n_M \rangle,
\end{equation}
with $c_{n_1,\dots,n_M}$ denoting the expansion coefficients. Thus, for the four-well system considered here we have
\begin{equation}
|\bm{\psi}, \N \rangle = \sum_{n_1,n_2,n_3,n_4}  c_{n_1,n_2,n_3,n_4} |n_1,n_2,n_3,n_4 \rangle.
\label{eq:allgFockZustand}
\end{equation}

The number of expansion coefficients $c_{n_1,n_2,n_3,n_4}$ is equal to the corresponding dimension $D$ of the Hilbert space. By contrast, there are just five real constraints \eref{eq:Bed1u2}--\eref{eq:Bed4u5}, which is not sufficient to determine the state uniquely. Thus, to find a suitable initial state we start with a \emph{pure} state, in particular the {\PT}-symmetric MF state \eref{eq:eigen}.

A pure condensate in an $M$-well system is a product state of identical single-particle states, where the total particle number in the system is fixed. This product state can be written as \cite{Dast2014}
\begin{equation}
|\bm{\psi}, \N \rangle =\sum_{n_1,\dots,n_M}   \underbrace{\sqrt{\frac{\N!}{n_1!\dots n_M!}} \psi_1^{n_1} \dots \psi_M^{n_M}}_{=c_{n_1,\dots,n_M}} |n_1, \dots, n_M \rangle, \label{eq:reinFockDarstellung_M}
\end{equation}
with $\psi_1, \dots, \psi_M$ the MF coefficients in well $i \in \{1, \dots, M \}$. Therefore, the {\PT}-symmetric state \eref{eq:psi1psi4} of the four-well system in Fock space can be written as
\begin{equation}
|\bm{\psi}, \N \rangle = \sum_{n_1,n_2,n_3,n_4}  \underbrace{\sqrt{\frac{\N!}{n_1!n_2!n_3!n_4!}} \psi_1^{n_1} \psi_2^{n_2} \psi_3^{n_3} \psi_4^{n_4}}_{=c_{n_1,n_2,n_3,n_4}} |n_1, \dots, n_4 \rangle,
\label{eq:reinFockDarstellung_M4}
\end{equation}
with $\psi_1, \dots, \psi_4$ the MF coefficients given in \eref{eq:psi1psi4}. This state is a pure state, hence the solutions \eref{Lsg_epsilon1} and \eref{Lsg_epsilon4} are not applicable, because the equations in the system \eref{LGS_epsilon_a} and \eref{LGS_epsilon_b} are linearly dependent in this case and there is not a single unique solution for the onsite energies.

To obtain a mixed state, the product state \eref{eq:reinFockDarstellung_M4} is perturbed by deflecting it with normally-distributed random numbers. We use a normal distribution with mean $m=1$ and variance $d$, which is freely selectable and a measure for the strength of the deflection. The state now for $M$ wells has the form
\begin{equation}
|\bm{\psi}, \N \rangle = \sum_{n_1,\dots,n_M} z_{n_1,\dots,n_M} \sqrt{\frac{\N!}{n_1!\dots n_M!}} \psi_1^{n_1} \dots \psi_M^{n_M}  |n_1,\dots,n_M \rangle,
\label{eq:reinFockDarstellung_M_z}
\end{equation}
with the random number $z_{n_1,\dots,n_M}$ belonging to the basis vector $|n_1, \dots, n_M \rangle $. For $M=4$ wells we have
\begin{equation}
\hspace{-1.3cm} |\bm{\psi}, \N \rangle = \sum_{n_1,n_2,n_3,n_4} z_{n_1,n_2,n_3,n_4} \sqrt{\frac{\N!}{n_1!n_2!n_3!n_4!}} \psi_1^{n_1} \psi_2^{n_2} \psi_3^{n_3} \psi_4^{n_4}  |n_1,n_2,n_3,n_4 \rangle.
\label{eq:reinFockDarstellung_M4_z}
\end{equation}

The state \eref{eq:reinFockDarstellung_M4_z} still has $D$ degrees of freedom, which now will be used to fulfill the requirements \eref{eq:Bed1u2}--\eref{eq:Bed4u5}. Since the random numbers only describe a perturbation of the pure state, they can be adjusted retrospectively in such a way that the resulting state fulfills the requested requirements. Numerically, we vary the  coefficients $z_{n_1,n_2,n_3,n_4}$ by using a minimization technique.

Note that basically more than just five requirements could be fulfilled by using this ansatz, because there are $D$ degrees of freedom available. This point is further discussed in \sref{sec:section5}.

\section{Numerical results}
\label{sec:section5}

In this section we present and discuss our numerical results. The calculations were performed using the Bose-Hubbard Hamiltonian \eref{eq:BHM_Ham} with the numerical exact methods from Refs.~\cite{Zhang_exakt_diag, Dizdarevic2018}. The same methods can be used to evaluate the elements of the singple-particle density matrices \eref{eq:sigma_kl} and the two-particle density matrices \eref{eq:sigma_klmn}. We show that by using the time dependences \eref{eq:tunnelJ12}--\eref{eq:tunnelJ34} and \eref{Lsg_epsilon1}--\eref{Lsg_epsilon4} of the control parameters, obtained by the numerical evaluation of equations \eref{eq:sigma_kl} and \eref{eq:sigma_klmn} using the exact time evolution, we are able to adjust the dynamics of the {\PT}-symmetric state \eref{eq:eigen} of the two-mode model \eref{eq:GPE_twomode} in the first order of the four-well many-body system. To achieve this, it is furthermore important to use appropriate initial states, which we constructed as shown in \sref{sec:section4}. We are especially interested in the behaviour of the system in the limit of low particle numbers, therefore we show, in an exemplary way, the dynamics of a system with a total particle number $\N = 22$. To allow for easy comparison with results from previous papers \cite{Graefe2012, Dast2014, Dast2016, Kreibich2013, Graefe2010}, without loss of generality we choose $J=1$. We first analyse the dynamics of first-order moments and then discuss the dynamics of the second-order moments.

\paragraph{Dynamics of first-order moments}

\begin{figure}
%\centering
\includegraphics[width=\textwidth]{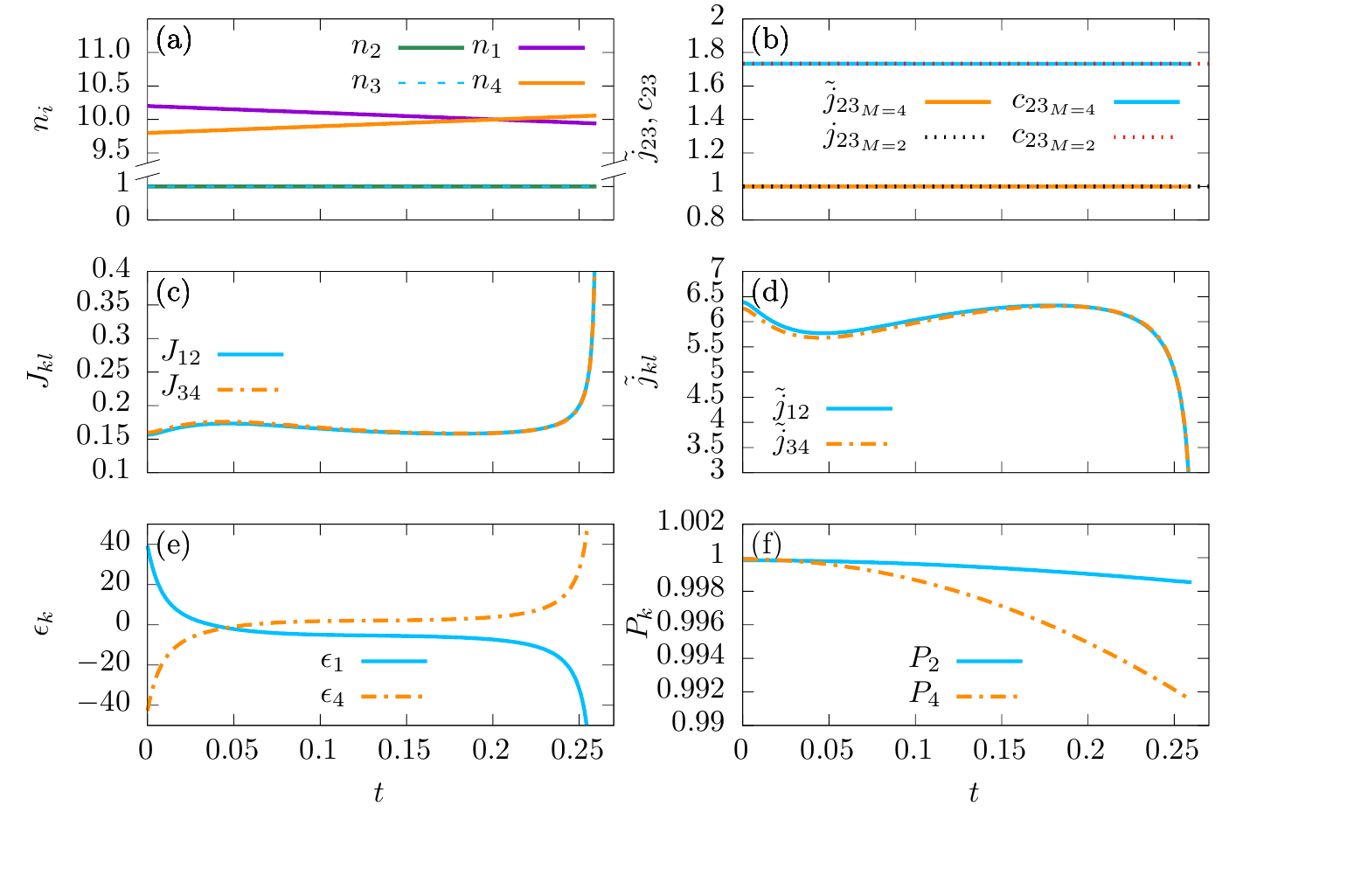}
\caption{\label{fig:ersteMomente1}
Dynamics of a mixed initial state of the form \eref{eq:reinFockDarstellung_M4_z} in the four-well many-body system with a total particle number of $\N=22$, calculated with the Bose-Hubbard model and the time dependences \eref{eq:tunnelJ12}--\eref{eq:tunnelJ34} and \eref{Lsg_epsilon1}--\eref{Lsg_epsilon4} of the control parameters for the system parameters $\gamma=0.5$ and $U=0.1$ and a variance of $d=0.008$ for perturbing the pure state \eref{eq:reinFockDarstellung_M4}. The time dependences of the (a) occupation numbers $n_i$, (b) elements $\tilde{j}_{23}$ and $c_{23}$, (c) tunnelling rates, (d) elements $\tilde{j}_{12}$ and $\tilde{j}_{34}$, (e) onsite energies and (f) purity are plotted. The dynamics of the first order in the many-body system is equivalent to the dynamics in the two-mode system.}
\end{figure}

In \fref{fig:ersteMomente1} the dynamics of (a) the occupation numbers in well $i$, (b) the elements $\tilde{j}_{23}$ and $c_{23}$, (c) the tunnelling rates, (d) the elements $\tilde{j}_{12}$ and $\tilde{j}_{34}$, (e) the onsite energies, and (f) the purity are depicted. We chose $\gamma = 0.5$ and $U=0.1$ for the system parameters and perturbed the pure MF state \eref{eq:reinFockDarstellung_M4} with random numbers, which are normally distributed with mean $m=1$ and variance $d=0.008$. Our numerical results in \fref{fig:ersteMomente1}(a--b) clearly demonstrate that we are able to adjust the dynamics of the {\PT}-symmetric state \eref{eq:eigen} in the first order of the many-body dynamics of the four-well many-body system. We will discuss this fact in the following.

In \fref{fig:ersteMomente1}(a) the dynamics of the occupation numbers $n_i$, $i \in \{1,\dots, 4\}$ is depicted. Obviously the number of particles in the inner wells behave exactly as known from the two-mode model. We can achieve constant particle numbers in time with $n_2$ and $n_3$ being equal. The reservoir wells are emptied and filled linearly. This behaviour of the reservoir wells has already been observed in the MF limit \cite{Kreibich2013} and can be explained as follows. Since the whole system, i.e.\ inner and reservoir wells, is closed and the population in the inner wells should be stationary, an equal and steady particle current is needed through all wells. Therefore, the reservoir wells are emptied and filled linearly for the in-coupling and out-coupling, respectively. Thus as regards the dynamics of the occupation numbers $n_2$ and $n_3$ our approach produces the desired result.

\Fref{fig:ersteMomente1}(b) displays the time dependence of the reduced current density $\tilde{j}_{23}$ and the correlation $c_{23}$. The solid lines show the behaviour of the many-body system, denoted with the index $M=4$, whereas the dynamics of two-mode system is depicted using dotted lines and denoted with the index $M=2$. As can be clearly seen both systems show exactly the same dynamics of the plotted quantities. Combined with the results from \fref{fig:ersteMomente1}(a) this demonstrates  that our method is suitable for adjusting the dynamics of the open system in the first order of the four-well many-body system. By using mixed states it is possible to keep $n_2$, $n_3$, $\tilde{j}_{23}$ and $c_{23}$ constant in time. Consequently, all of the moments of first order in the many-body system show the same behaviour as those of the two-mode model. This is a result that clearly goes beyond \cite{Dizdarevic2018}, since we are able to achieve the stationarity of all four quantities of first order, which is not possible by using pure states.

As can be seen in \fref{fig:ersteMomente1}(c), the tunnelling rates $J_{12}$ and $J_{34}$ diverge after a certain time. This is the point in time when the system collapses, which means the breakdown of the system. The reason of this behaviour is the dynamical development of the reduced density currents $\tilde{j}_{12}$ and $\tilde{j}_{34}$, which is depicted in \fref{fig:ersteMomente1}(d). At the collapse point in time both of the values decrease to zero.  Since $\tilde{j}_{12}$ and $\tilde{j}_{34}$ are included in the calculation of the tunnelling rates according to \eref{eq:tunnelJ12}--\eref{eq:tunnelJ34}, i.e.\ as denominators, $J_{12}$ and $J_{34}$ diverge. For the same reason also the onsite energies $\epsilon_1$ and $\epsilon_4$ diverge at this point in time, as can be seen in \fref{fig:ersteMomente1}(e), since the values  $\tilde{j}_{12}$ and $\tilde{j}_{34}$ in \eref{Lsg_epsilon1} and \eref{Lsg_epsilon4} also occur in the denominators. This behaviour is a consequence of the limited number of control parameters, which allows only for the control of the single-particle dynamics in the inner wells. The influence of the higher-order dynamics onto the first order grows rapidly due to decoherence as the system evolves. To compensate such perturbations of the first-order dynamics, the control parameters are increased, which, however, does not directly influence the higher orders and thus cannot counter its effects specifically. This means that we cannot sustain the states for arbitrary times, but nevertheless for a finite time interval.

It should be noted that the time intervals for the many-body systems presented in this paper are short in comparison with those studied in the mean-field limit (cf.\ Refs.~\cite{Kreibich2013,Kreibich2014}). An estimate of the unit of time is given in Ref.~\cite{Kreibich2013}, that is, $t_l \approx 5.47\,\mathrm{ms}$, for a generic experimental realization using $^{87}$Rb atoms. In such an experiment stationary states could thus be realized for a few milliseconds. The specific value, however, sensitively depends on the structure of the state and there is no simple relation to the mean and the variance of the random numbers used to generate it.

Following Dast \etal \cite{Dast2016a}, we define the purity $P$ of a condensate in an $M$-well system as
\begin{equation}
P_M = \frac{M \tr \left( \sigma_{red} \cdot \sigma_{red} \right) - 1}{M-1}.
\end{equation}
This is a scaled definition of the purity, where $P_M$ can only take values between zero and one, while $P_M=1$ means that we have a perfectly pure condensate, i.e.\ the system is in a product state, and $P_M=0$ describes a maximally impure BEC. The index refers to the number of wells considered.

The time evolutions of the purity $P_4$ of the four-well system and the purity $P_2$ related to the embedded system are shown in \fref{fig:ersteMomente1}(f). In good approximation $P_2$ is constant in time. Since we used normally-distributed random numbers of a relatively small variance to perturb the pure state, both $P_2$ and $P_4$ are close to one, which means that the constructed state is not far away from a perfectly pure state. We see that already a small perturbation of the pure MF state \eref{eq:reinFockDarstellung_M4} is sufficient to realize the dynamics of the two-mode model in the first order of the many-body system. This means, that even for small perturbations of the pure state the system of equations \eref{LGS_epsilon_a} and \eref{LGS_epsilon_b} has the single unique solution \eref{Lsg_epsilon1}--\eref{Lsg_epsilon4} and the method presented here yields the requested behaviour, and it is absolutely not necessary to use a more mixed state.

\paragraph{Dynamics of second-order moments}

We now discuss the behaviour of the dynamics of the second-order moments $\sigma_{klmn}$.

In the two-mode model, the second-order moments can be obtained by using the pure state \eref{eq:reinFockDarstellung_M} and are given by
\begin{equation}
\sigma_{klmn}(t) = \N \left( \N - 1 \right) \psi_k^*(t) \psi_l(t) \psi_m^*(t) \psi_n(t) + \N \delta_{lm} \psi_k^*(t)  \psi_n(t)
\label{eq:Anfangswert_sigma_klmn}
\end{equation}
with $\delta_{lm}$ denoting the Kronecker delta. Inserting the {\PT}-symmetric eigenstate \eref{eq:eigen} yields all second-order elements in the two-mode model to be constant in time. Note that in the MF the moments of all orders of the hierarchy are time-independent.

Not all of the elements of second-order are independent of each other. Using the commutation relations of creation and annihilation operators for a multiple-boson system symmetries can be found, therefore the number of independent elements can be significantly reduced. Of special interest is the behaviour of the elements $\sigma_{klmn}$ with $k,l,m,n \in \{2,3\}$. For four wells, the number of such elements can be reduced to nine.

We do not show the dynamics of the second-order moments here explicitly, but rather explain why mixed initial states in the four-well many-body system constructed as described in \sref{sec:section4} do not automatically show the same second-order dynamics as \eref{eq:eigen} in the two-mode model. There are two points to be considered:

The first point is that we did not set conditions for initial values of the second-order moments, i.e.\ the values at $t=0$ in the two-mode model and in the many-body system generally differ from each other. Nevertheless, our numerical calculations show \cite{Mathea_master} that they are of the same magnitude. This implies that in principle the values of the two-mode model can be adjusted in the many-body system at $t=0$. However, doing this for our four-well system nevertheless would not be worthwhile, which becomes clear with our second point.

The second point is the following: In contrast to the MF case, in the many-body description four real control parameters are not enough to generate the same dynamics in the many-body system as in the two-mode model beyond the first order. Assuming that one has adjusted the correct values for the second-order moments in the four-well system, one would not gain anything. This is due to the number of control parameters on hand in the four-well system. There are only four real parameters which can be varied in time to adjust the behaviour present in the two-mode model. They were used to make the values $n_2$, $n_3$, $\tilde{j}_{23}$ and $c_{23}$ fit in both systems. Thus, in the four-well many-body system there are no parameters left to control the behaviour of higher-order elements, therefore the dynamics of second-order elements is not the same as in the two-mode model. Thus, even if we adjusted the correct initial values in second order, we would not be able to keep these values constant.

The four-well system is a minimal setup to simulate the dynamical properties of the non-Hermitian two-mode model in a larger Hermitian system \cite{Kreibich2013}. The realization of the first-order behaviour of the {\PT}-symmetric state \eref{eq:eigen} is the maximum of what is possible in the four-well many-body system. To also adjust higher-order dynamics in a many-body system, one could add more reservoir wells to have more control parameters at hand. The method to find an appropriate initial state presented here could also be used to fulfill further conditions.

\section{Summary and conclusion}
\label{sec:summary}

We presented a method to adjust the first-order dynamics in the two inner wells of a four-well system to be identical to the dynamics of the {\PT}-symmetric eigenstate of the two-mode model describing BECs with balanced gain and loss. For this purpose, we used time-dependent control parameters in the many-body system. The time dependence of the tunnelling rate is given by \eref{eq:tunnelJ12}--\eref{eq:tunnelJ34}, the one of the onsite energies was chosen according to \eref{Lsg_epsilon1} and \eref{Lsg_epsilon4}. It has been shown that initial states have to fulfill the five real requirements \eref{eq:Bed1u2}--\eref{eq:Bed4u5}. We showed one possible way to construct mixed initial states for the many-body dynamics based on MF states fulfilling the five constraints.

Our numerical results clearly demonstrate that the method presented is suitable to achieve a stationary behaviour of the occupation numbers $n_2$ and $n_3$ in the inner wells, the reduced current density $\tilde{j}_{23}$ and the correlation $c_{23}$, which fully characterize the system in first order. This result clearly goes beyond previous work \cite{Kreibich2013, Dizdarevic2018}, in which this was realized either in the MF limit or only the stationarity of $n_2$, $n_3$ and $\tilde{j}_{23}$ could be achieved. By contrast, we were able to adjust the {\PT}-symmetric state's complete first-order dynamics in the many-body system.

The behaviour of the second-order elements and higher-order elements in the many-body system differs from that in the two-mode model. Since the four-well system only holds four control parameters, the maximum that is possible is to adjust the dynamics to the first order, which is characterized by four values. To also control higher-order behaviour, a higher number of control parameters would be necessary. This could be achieved, for example, by adding more reservoir wells. By using the method for choosing an appropriate initial state presented here it would be possible also to fulfill further conditions resulting from higher orders. Combined with suitable time dependences of a greater number of control parameters one could control the behaviour of higher-order dynamics in the many-body system as well.

% -----------------------------------------------------------------------------
\section*{References}

%\bibliographystyle{unsrt}
%\bibliography{paper}

\end{document}